\documentclass[RNAAS]{aastex62}

\begin{document}

\title{Multi-wavelength view of the M87 black hole captured by Event Horizon Telescope}

\correspondingauthor{Giacomo Principe}
\email{giacomo.principe@infn.it}

\author[0000-0003-0406-7387]{G. Principe}
\affiliation{Dipartimento di Fisica, Universit\'a di Trieste, I-34127 Trieste, Italy}
\affiliation{Istituto Nazionale di Fisica Nucleare, Sezione di Trieste, I-34127 Trieste, Italy}
\affiliation{INAF Istituto di Radioastronomia, Via P. Gobetti, 101,  I-40129 Bologna, Italy}

\author[0000-0000-0000-0000]{M. Lucchini}
\affiliation{MIT Kavli Institute for Astrophysics and Space Research, MIT, 70 Vassar Street, Cambridge, MA 02139, USA}
\affiliation{API--Anton Pannekoek Institute for Astronomy, University of Amsterdam, Science Park 904, 1098 XH Amsterdam, The Netherlands}

\author{on behalf of Event Horizon Telescope (EHT) Multi-Wavelength Science Working Group}
\author{EHT Collaboration}
\author{Fermi-LAT Collaboration}
\author{H.E.S.S. Collaboration}
\author{MAGIC Collaboration}
\author{VERITAS Collaboration}
\author{EAVN Collaboration}


\section{Abstract} 
In 2017, the first image of the center of the M87 galaxy was captured by the Event Horizon Telescope (EHT). It revealed a ring morphology and a size consistent with  theoretical expectations for the light pattern around a weakly accreting supermassive black hole of $\sim$6.5 billion solar masses. In parallel to the EHT measurements, an extensive multi-wavelength campaign with ground- and space-based facilities from radio all the way up to the TeV energy range was organized. In this note we will give an overview of the results from this campaign. M87 was found to be in a historically low state. At X-ray energies the emission from the core dominates over HST-1. We present the most complete simultaneous, multi-wavelength spectrum of the active nucleus to date. 
\\
\section{Introduction}

M87 is the most prominent elliptical galaxy within the Virgo Cluster, located just $16.8\pm 0.8$\,Mpc away \citep{2009ApJ...694..556B,2010A&A...524A..71B}. 
In addition of being among the closest active galactic nuclei (AGN), M87 also hosts the first example of an extragalactic jet to have been detected by astronomers \citep{1918PLicO..13....9C}.
This famous one-sided jet has been well-studied in almost every waveband from radio (down to sub-parsec scales) to gamma rays \citep[e.g.,][]{1989ApJ...336..112R,1999ApJ...520..621B,2002ApJ...564..683M,2009ApJ...707...55A}.
Despite the many studies that have been performed, the conditions under which such jets are launched, as well as the origin of gamma-ray emission in radio galaxies are two of the most enduring questions in astrophysics today \citep[see e.g.,][]{1977MNRAS.179..433B,1982MNRAS.199..883B}. 

In 2017, the first direct image of an super-massive black hole (SMBH) “shadow” in the center of M87 galaxy was captured by the Event Horizon Telescope (EHT) \citep{2019ApJ...875L...1E}. 
It revealed a ring morphology and a size consistent with theoretical expectations for the light pattern around a weakly accreting supermassive black hole of $\sim$6.5 billion solar masses. The mass of the SMBH was derived by comparing the size of blurry ring that was measured to that predicted by models \citep{2019ApJ...875L...1E}. 

To test the newest generation of models, as well as to investigate the origin of the high-energy emission, it is important to have extensive, quasi-simultaneous multi-wavelength (MWL) monitoring of AGNs, providing both spectral and imaging data over a wide range of physical scales.
We present here an overview of the results from the 2017 EHT campaign on M87, combining very long baseline interferometry (VLBI) imaging and spectral index maps at longer wavelengths, with spectral data from submillimeter (submm) to TeV gamma rays (covering more than 17 decades in frequency).
\\
\section{Observational campaign and results}

In parallel to the EHT measurements, an extensive multi-wavelength campaign with ground- and space-based facilities from radio to TeV gamma rays was organized. This provided the most extensive, quasi-simultaneous, broadband spectrum of M87 yet taken, together with the highest ever resolution mm-VLBI images using the EHT from its 2017 April campaign.

\begin{figure}[h!]
\begin{center}
\includegraphics[scale=0.6, trim= 2cm 0cm 2cm 0cm]{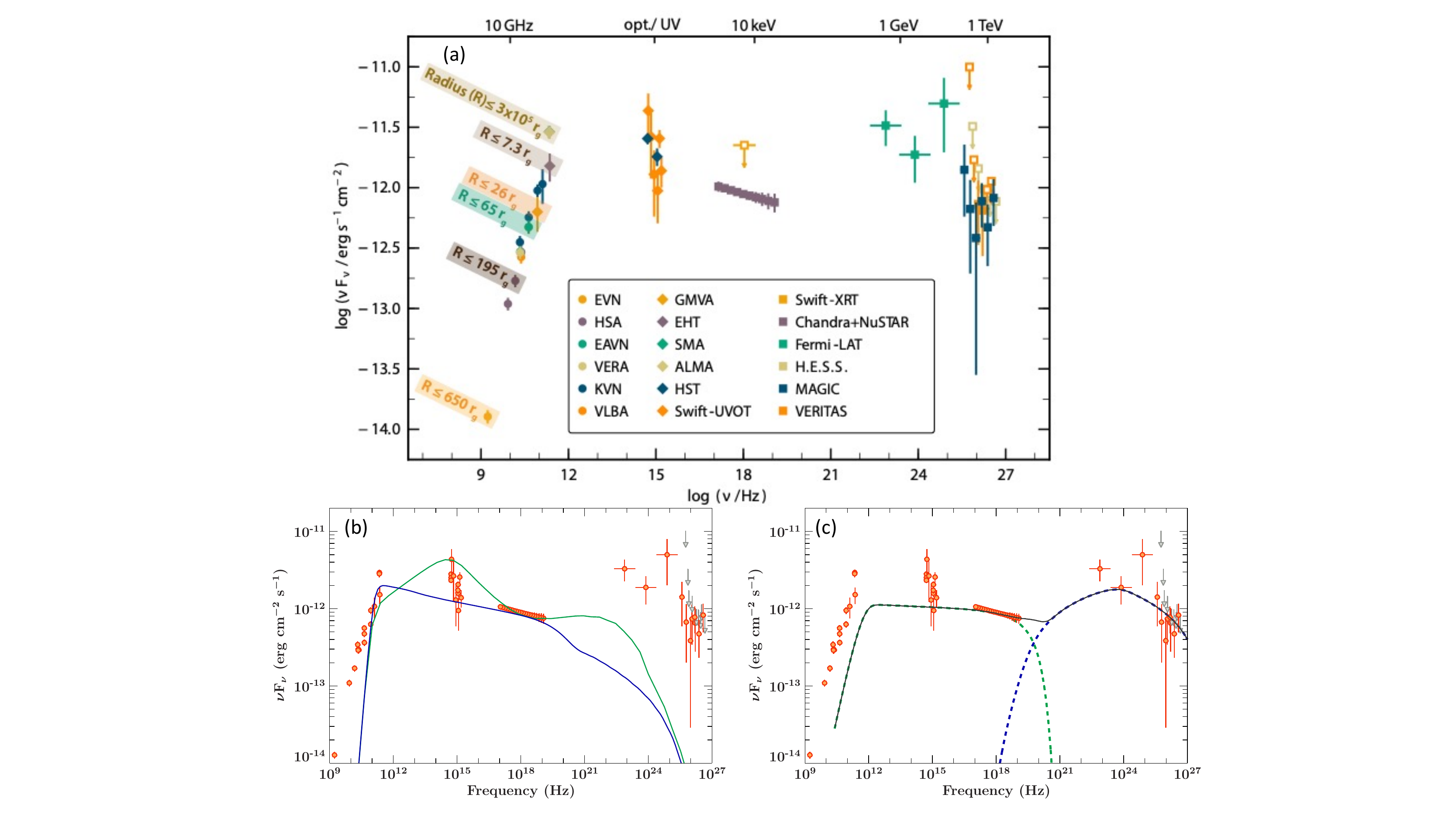}
\caption{Top: Observed broadband SED of M87 quasi-simultaneous with the EHT campaign in April 2017 with fluxes measured by various instruments highlighted with different colors and markers. For the mm-radio VLBI, the upper limits on emission size for several representative frequencies are labelled to clarify the size-scale. An illustration of the resolved flux differences depending on spatial resolution is shown by the comparison of the differing EHT and ALMA-only 230 GHz fluxes and size limits. (b): SED fit focusing on the EHT data. Blue and green lines display the resulting SEDs for models including radiative cooling on the initial single power-law eDF and a broken power-law, respectively.
(c) SED with model 2 fit focusing on the higher energy data.
The figures are taken from \citet{2021ApJ...911L..11E}. \label{fig:1}}
\end{center}
\end{figure}

M87 was found to be in a historically low state, both for its core and for the nearest knot HST-1. In the X-ray range, the emission from the core dominates over HST-1. This provides the ideal observing conditions for a multiwavelength campaign combining data over a large range of spatial resolution, since M87’s core was dominating the total flux in the radio through X-ray bands. This made it possible to combine core flux constraints with the more spatially precise VLBI data.
Figure \ref{fig:1}(a) shows the broadband SED of M87 over a the range between a frequency of $\simeq\!1$\,GHz and photon energy of $\simeq\!1$\,TeV. The emission measured by the instruments contributing to this broadband SED spans a factor of $\gtrsim10^8$ in angular scale: from the effective spatial resolution of the EHT, $\simeq\!20$\,$\mu$as \citep{2019ApJ...875L...4E}, to the resolving power of \textit{Fermi}-LAT in low-energy $\gamma$-rays, $\simeq\!2^{\circ}$ \citep{2018A&A...618A..22P}.
In Figure \ref{fig:1} we model the SED focusing on the EHT data. At $\gamma$-ray energy bands, one can see the SSC emission components although they underestimate
the observed gamma-ray flux density.
In Figure \ref{fig:1} (c) the SED is modelled focusing on the higher energy data. Without a strong prior constraint on the size, as in the previous case, we found that most model parameters are highly degenerate and cannot be constrained satisfactorily by our fit.

\section{Conclusions}
The goal of this extensive, quasi-simultaneous observational campaign was to provide this important legacy data set, which will serve as a resource for the entire community, to enable the best possible modelling outcomes and a reference point for theory. Based on simple single-zone models, it appears that M87’s complex, broadband spectral energy distribution cannot be modeled by a single zone. The stratified nature of the radio through millimeter bands is clearly indicated by the high dynamic-range images that place significant size constraints on the emitting geometry per frequency band. 
Furthermore it is not yet clear where the gamma rays originate, but it is possible to rule out that they are originated within the EHT region by leptonic processes.
Although a $\sim$100 times larger region can provide the measured gamma-ray emission, it would, however, require an uncomfortably high particle domination compared to the energy stored in the magnetic fields, which may be inconsistent with the observed jet velocity and parabolic geometrical profile.
Further information on the results and a description of the analysis performed for this campaign, as well as the multi-wavelength data as a legacy data repository, can be found in \citet{2021ApJ...911L..11E}.

\acknowledgments
The support given by the LAT AGN Working Group coordinators to review this note, is gratefully acknowledged. We refer to the related publication \citep{2021ApJ...911L..11E} for the complete acknowledgments.

\end{document}